# Configured Grant for Semi-Deterministic Traffic for Ultra-Reliable and Low Latency Communications

Bikramjit Singh
*Ericsson Research*
Jorvas, Finland
bikramjit.b.singh@ericsson.com

Majid Gerami
*Ericsson Research*
Lund, Sweden
majid.gerami@ericsson.com

*Abstract*—**Configured Grant-based allocation has been adopted in New Radio 3rd Generation Partnership Project Release 16. This scheme is beneficial in supporting Ultra-Reliable and Low Latency Communication for industrial communication, a key Fifth Generation mobile communication usage scenario. This scheduling mechanism enables a user with a periodic traffic to transmit its data readily and bypasses the signaling entailed to scheduling requests and scheduling grants; and provides low latency access. To facilitate ultra-reliable communication, the scheduling mechanism can allow the user to transmit redundant transmissions at consecutive repetition occasions in a pre-defined period. However, for the user with semi-deterministic traffic, the reliability and latency performance with Configured Grant-based allocation deteriorates. This can be due to, e.g., late data arrival in the buffer, and the user unable to transmit its repetitions, which leads to reliability degradation. To improve the Configured Grant reliability performance with semi-deterministic traffic, we consider various allocation designs utilizing, e.g., additional unlicensed spectrum, or flexible transmission in a Configured Grant period, or allowing time-gaps between the repetitions, etc. These enhancements could be a stepping-stone for Sixth Generation Configured Grant models.**

*Keywords—Configured Grant, Ultra-Reliable And Low Latency Communications*

## I. INTRODUCTION

The next generation of mobile communication systems known as Fifth Generation (5G) will be launched around the world in the year 2020. Ultra-reliable and low latency communications (URLLC) is one of the key usage scenarios that is targeted for beyond International Mobile Telecommunications-Advanced services [1]. This enables real-time control and automation of dynamic processes for vertical applications such as factory automation, transport industry (e.g., remote driving), smart grids and electrical power distribution. These services may require reliability of an order of 99.9999 % and low radio latency down to 0.5 ms. To support such services with extreme requirements, the 5G New Radio (NR) networks are being architected.

To fulfill the stringent requirements on low-latency access for URLLC services, especially for an uplink transmission, it is important to eliminate the delay for Scheduling Request (SR) from a User Equipment (UE) to the next generation Node B (gNB) and the corresponding delay for scheduling grant from the gNB to the UE [2]. By excluding these signaling, it further improves the reliability of an uplink transmission by discounting the probability failure of SR transmitted over uplink control channel and grant over downlink control channel. The data is transmitted readily on the allocated resource. Therefore, instead of scheduling a transmission on dynamic basis, the radio resources are allocated to the UE in a periodic manner approximately mirroring to its periodic traffic needs. This scheduling feature of uplink transmission without requiring SR and grant is denoted as Configured Grant (CG) in 5G NR.

To improve the CG reliability, the allocation can be supported with $K$ consecutive redundant allocations or Transmission Occasions (TOs) in a pre-defined period. The length of the period can be established as per the latency budget of the transmission. These TOs are assigned with a definitive Redundancy Version (RV) pattern. RV defines the decoding capability of a repetition, e.g., RV 0 repetition is a self-decodable, RV 3 is almost decodable, and the repetitions with RV 1 and 2 require repetitions with other RVs for their decoding. NR Release 16 has specified three RV patterns for the repetitions in CG, i.e., (0,0,0,0), (0,3,0,3) and (0,2,3,1).

In URLLC industrial use-cases, e.g., where a sensor reports an alarm events in a factory, the event arrival can be sporadic or semi-deterministic. In such scenarios where the UE scheduled with CG, if it misses TO(s) in a period, then it either should transmit lesser than $K$ repetitions or should wait for until next period to transmit all the repetitions. Further, the CG transmissions may also get impacted due to time-division duplex patterns. The repetitions are discounted if the resource allocation happens to be a downlink resource. In some scenarios, the resource can be utilized by high priority traffic and the CG transmissions can get preempted. Resultant, this may reduce reliability or increase the latency in case the UE postpones its transmissions to the next CG period.

In NR Release 16, to reduce the latency, the UE can be configured with multiple configuration where each configuration has different starting offset. Hence, in case of fluctuating traffic, the UE picks the nearest available CG. The use of multiple configurations in NR is not limited to the above application of reducing alignment delay. In addition, if the UE is expected to carry different traffic, then multiple CGs can be exploited where each configuration can have different parameters in periodicity, priority, reliability and so on. However, using multiple CGs for reducing latency can be spectrum inefficient, as only one configuration is utilized at a time.

In summary, the current 3rd Generation Partnership Project (3GPP) CG model is not fully equipped to cater semi-deterministic traffic. Hence, in this extended abstract, we look to explore briefly some notable CG strategies that can provide better reliability and latency performance for such traffic and can be considered for future releases towards Sixth Generation.

## II. CONFIGURED GRANT FOR SEMI-DTEREMINITIC TRAFFIC

Network can implement a suitable CG strategy depending on the traffic, e.g., UE population, traffic arrival, to achieve the URLLC reliability target within the latency budget, and simultaneously reduce the resource usage.

### A. Flexible K repetitions offset

3GPP allows two mechanism for repetitions transmission on $K$ consecutive TOs. In one scheme, the UE always starts with the first TO. If the data comes after the first TO, the UE does not transmit at all. Therefore, either the UE transmits all its $K$ repetitions or none. In another mechanism, the UE can start repetitions transmission from any TO configured with RV 0 in the TOs' RV pattern and continue with the following repetitions on the remaining TOs as per RV pattern. Here, the UE could accommodate the late traffic arrival by transmitting lesser than $K$ repetitions. However, both schemes degrade with semi-deterministic arrival, especially in the first scheme, where it witnesses complete loss in reliability for the late arrivals.

To enhance reliability for the semi-deterministic traffic, $T$ TOs instead of $K$ TOs in a CG period can be allocated such that $T > K$. This allows the UE to perform all $K$ repetitions if the traffic arrives anywhere in the beginning of $T - K$ TOs, or with partial repetitions in the later half. The network can estimate the resource $T$ TOs such that $T = \min_{P_r(\gamma, r) \geq P_r^T} r$ where $P_r(\gamma, r)$ is the transmission reliability given $r$ TOs are allocated in a CG period and $\gamma$ is mean-arrival time, and $P_r^T$ is the reliability target.

In the 3GPP model, the TOs have a fixed RV pattern, and this forces the UE to always begin with repetition over a TO configured with RV 0 which is fully decodable RV. However, in the proposed scheme, the transmitted $K$ repetitions follow the pattern instead the allocated TOs. Whenever a gNB detects $K$ repetitions over $T$ TOs, it decodes them with the same defined pattern. Hence the scheme does not force the UE to begin with specific TOs which limits the reliability performance for the semi-deterministic traffic unlike in the current 3GPP model. Alternatively, gNB employs blind decoding of the first repetition as it may happen anywhere on the available TOs. However, the CG resource is allocated in a dedicated manner, therefore, recognizing the UE profile of the first repetition should not be a problem for the gNB.

### B. Time-gap in K repetitions

3GPP model stipulates the allocation of $K$ TOs in a consecutive manner. This is adequate as it provides low latency access. However, for semi-deterministic traffic, this can harm as these TOs wasted fast if the data arrives late in a period, and the resultant repetition count would be less, leading to poor transmission reliability. To reduce the wastage or improve the reliability performance, a time-gap can be introduced in the $K$ TOs which can act as a deterrence to the fluctuating arrival. This can be done if the time-budget for the URLLC transmission does not violate. Although, some TOs still be wasted but the proportion would be less in comparison to consecutive allocations for the same amount of TOs.

### C. Assistsive shared spectrum usage

For a semi-deterministic traffic, to allow the UE to transmit all $K$ repetitions in a period, the gNB must over allocate $T$ TOs in period such that $T > K$. This will improve reliability at the expense of extra licensed spectrum usage. Given the data arrival has a certain degree of randomness, hence in those scenarios, where the data arrives in the beginning of a period, the UE utilizes first $K$ TOs, and the remaining $T - K$ TOs remains unutilized as the UE's needs have already fulfilled. These unutilized spectrum resources can be utilized for other needs if gNB acts fast, e.g., by allocating resources to best-effort, streaming traffic. However, if the UE arrives late in the period, the initial TOs will be wasted. The resource wastage probability can be calculated as $p_o \times T + \sum_{i=1}^{T} p_i \times (i - 1)$ where $p_o$ is the probability of no arrival and $p_i$ is the probability of the data arrival in $i$-th TO.

To reduce the wastage, a shared spectrum can be allocated to assist CG. The shared spectrum can compose of licensed spectrum, e.g., in the form of a limited spectrum pool in [3] or an unlicensed spectrum, e.g., NR-Unlicensed (NR-U). Especially, in case of NR-U, there will be no added cost due to unlicensed spectrum usage, but time-budget may be required on sensing mechanisms, e.g., Listen-Before-Talk.

The shared spectrum allows the UE to transmit its remaining repetitions which it could not transmit in the CG period for whatsoever reasons. Therefore, in this, the UE is always allocated $K$ TOs in the CG period. Let us say, the UE transmits $K_{CG}$ repetitions on $K_{CG}$ TOs out of $K$ available TOs, where $K_{CG} < K$. To maintain reliability, the UE transmits remaining $K - K_{CG}$ repetitions in the shared spectrum within a specified latency budget. The UE may choose to transmit in a random manner in the shared resource unlike in CG resource. However, if the shared spectrum allows contention-based access [4] likewise in slotted-ALOHA, then the repetitions may collide with other transmissions. Hence, every repetition in the shared part has an additional associated failure probability due to the collision.

## III. FUTURE WORK

In this extended abstract, few strategies have been discussed for the advanced Configured Grant scheduling mechanism. In the future work, these strategies will be analyzed analytically, and reliability and latency performance will be verified using simulations.